\documentclass{aanew}
\usepackage{psfig}

\def\bron{XTE~J1908+094}
\def\ecs{erg~cm$^{-2}$s$^{-1}$}
\def\lum{erg~s$^{-1}$}

\begin{document}

\title{Broad-band X-ray measurements of the black hole candidate \bron}
\titlerunning{\bron}
\authorrunning{J.J.M. in 't Zand, J.M. Miller, T. Oosterbroek, A. Parmar}

\author{
J.J.M.~in~'t~Zand\inst{1,2}
\and J.M.~Miller\inst{3}
\and T. Oosterbroek\inst{4}
\and A.N.~Parmar\inst{4} 
}

\offprints{J.J.M. in 't Zand, email {\tt jeanz@sron.nl}}

\institute{     
		 Astronomical Institute, Utrecht University, P.O. Box 80000,
		 NL - 3508 TA Utrecht, the Netherlands
	  \and
		 SRON National Institute for Space Research, Sorbonnelaan 2,
		 NL - 3584 CA Utrecht, the Netherlands 
	  \and
		 Center for Space Research and Department of Physics,
                 Massachusetts Institute of Technology, Cambridge,
                 MA 02139-4307, U.S.A.
	  \and
		 Astrophysics Division, Research \& Scientific Support
                 Department, ESA, ESTEC SCI-SA, NL - 2200 AG Noordwijk, the
                 Netherlands
	}

\date{Received, accepted }

\abstract{ \bron\ is an X-ray transient that went into outburst in
February 2002.  After two months it reached a 2-250 keV peak flux of 1
to $2\times10^{-8}$~\ecs. Circumstantial evidence points to an
accreting galactic black hole as the origin of the the X-radiation:
pulsations nor thermonuclear flashes were detected that would identify
a neutron star and the spectrum was unusually hard for a neutron star
at the outburst onset. We report on BeppoSAX and RXTE All Sky Monitor
observations of the broad-band spectrum of \bron. The spectrum is
consistent with a model consisting of a Comptonization component by a
$\sim40$~keV plasma (between 2 and 60 keV this component can be
approximated by a power law with a photon index of 1.9 to 2.1), a
multicolor accretion disk blackbody component with a temperature just
below 1~keV and a broad emission line at about 6~keV.  The spectrum is
heavily absorbed by cold interstellar matter with an equivalent
hydrogen column density of $2.5\times10^{22}$~cm$^{-2}$, which makes
it difficult to study the black body component in detail. The black
body component exhibits strong evolution about 6 weeks into the
outburst. Two weeks later this is followed by a swift decay of the
power law component. The broadness of the 6 keV feature may be due to
relativistic broadening or Compton scattering of a narrow Fe-K line.
 }

\maketitle 

\section{Introduction}
\label{intro}

\bron\ was discovered serendipitously in 2--30 keV observations of
SGR~1900+14, at 24\arcmin\ distance, with the Proportional Counter
Array (PCA) on board RXTE on 2002 February 21 (Woods et al. 2002). The
source spectrum was consistent with an absorbed power law with a
photon index of $\Gamma=1.55$ and an equivalent hydrogen column
density of $N_{\rm H}=2.3\times10^{22}$~cm$^{-2}$. The power density
spectrum failed to show coherent pulsations and exhibited a flat
spectrum between 1~mHz and 0.1~Hz and a broken power law beyond that
up to 4~Hz, with a fractional rms amplitude of 43\% between 1~mHz and
4~Hz.

\bron\ was also strongly detected in a 15--250~keV serendipitous
observation of SGR~1900+14 on March 9.4--12.2 (UT) with the Phoswich
Detection System (PDS) on BeppoSAX (Feroci \& Reboa 2002). The spectrum
was seen to have a cutoff at about 100 keV. 

\bron\ was first imaged with the 2--10 keV Medium-Energy Concentrator
Spectrometer (MECS) on BeppoSAX on April 2.4-3.8 and an accurate
(20\arcsec\ error radius at 90\% confidence) was obtained (In~'t~Zand
et al. 2002a).  This enabled Rupen et al. (2002) to identify a radio
transient inside the MECS error circle at 4.86 and 8.46~GHz from
observations with the Very Large Array on March 21 and 22.  No obvious
variability nor extension was detected. The association of the radio
source with the X-ray source is confirmed by a short (1~ksec)
spatially high-resolution X-ray observation with the ACIS-S detector
on the {\em Chandra} X-ray observatory on April 15. We have analyzed
the wings of the heavily piled-up source image and find, if ones
assumes that the point spread function is symmetric, that the centroid
is at equatorial coordinates $\alpha_{2000.0}=19^{\rm h}08^{\rm
m}53\fs111$ and $\delta_{2000.0}=+9^\circ23\arcmin5\farcs57$ with an
uncertainty of 1\arcsec. The radio source is 0\farcs86 from this {\em
Chandra} position.

The hard spectrum and lack of pulsations provide evidence that we are
dealing with an X-ray binary in which the compact object is a stellar
black hole (BH; Woods et al. 2002). Proof of this needs to come from
the measuring the mass function through the radial velocity curve of
the companion. The prospects for this are not negligible despite the
large absorption; a near-infrared counterpart of $Ks=16.4$~mag has
been identified by Chaty \& Mignani (2002) whose location is
consistent with that of the radio counterpart. Searches in $R$ (Wagner
\& Starrfield 2002) and $I$ (Garnavich et al. 2002) were negative
with thresholds of 23 and 22 mag, respectively.

In this paper, we report on the follow-up observations of this source
with two of the four Narrow-Field Instruments (NFI) on BeppoSAX
(Boella et al. 1997a) during March 31 through April 3 that provide
unique information on the broad-band spectrum of \bron. This analysis
is supported with observations done with RXTE's All-Sky Monitor and
BeppoSAX' Wide Field Cameras.

\begin{figure}[t]
\psfig{figure=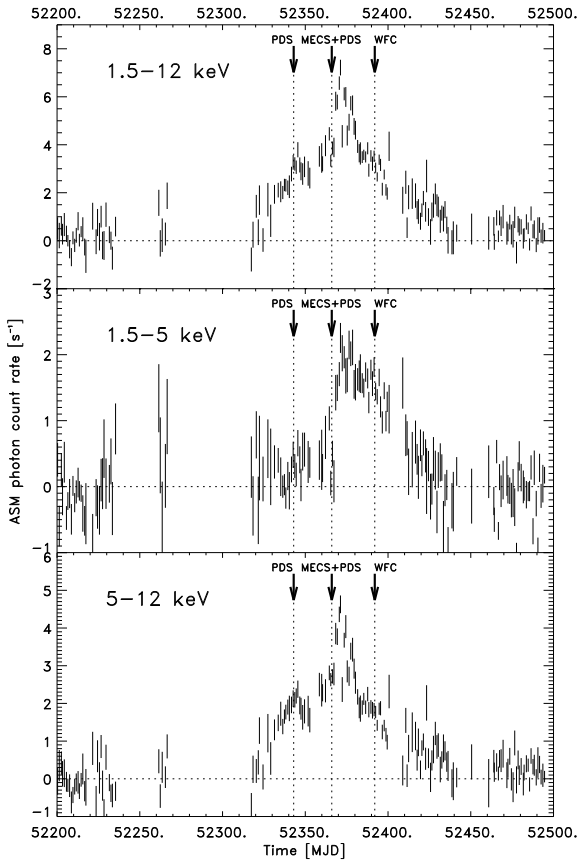,width=\columnwidth,clip=t}
\caption{One-day averages of \bron\ intensity as measured with ASM in
complete band (upper panel), and in two sub bands, updated to 2002 August
8.  The arrows and vertical dotted lines indicate the times of the
1st PDS observation (on March 9--12; arrow pointing to March 10), the
2nd PDS observation and the MECS observation (on March 31 to April 3;
arrow pointing to April 2) and the WFC observations (on April 27 and
29; arrow pointing to April 28).
\label{figasm}}
\end{figure}

\begin{figure}[t]
\psfig{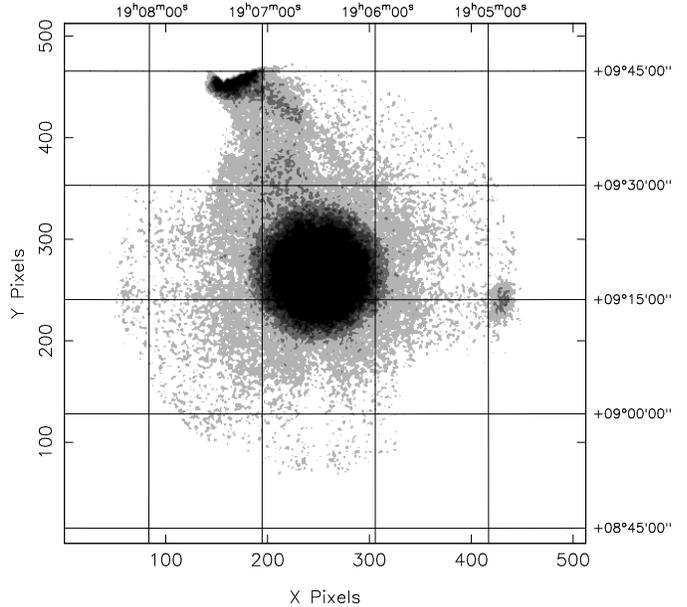}
\caption{Image obtained from the complete MECS observation in the full
bandpass (1.8--10.5~keV).  The image has been smoothed with a Gaussian
of standard deviation 8\arcsec.  Apart from \bron\ in the center, two
other sources are discernible: 4U~1907+097 (along the top edge, with a
point-spread function that has been cut out by an in-flight
calibration source holder) and SGR~1900+14 (weak source to the right).
\label{figmecsimage}}
\end{figure}

\section{Observations and data analysis}
\label{obs}

\subsection{ASM}

The RXTE All-Sky Monitor (ASM; Levine et al. 1996) measures the
1.5--12 keV flux of \bron\ up to 30 times a day during 90~s long
dwells with one of three Scanning Shadow Cameras. The flux is measured
in 3 channels. Figure~\ref{figasm} show the light curve derived by
binning the data in one-day intervals.  A first detection is apparent
on about 2002 February 15 (MJD~52320) at a level of about
1.5~c~s$^{-1}$ in the 1.5 to 12 keV band.  Subsequently the flux
increased for two months at which point it reached 7~c~s$^{-1}$ or
roughly one tenth the intensity of the Crab source.  The
dwell-by-dwell light curve does not reveal higher fluxes. The peak was
maintained for just a short time, perhaps a few days. After that the
source decayed over the course of about 2 months. There were no prior
detections in the ASM database that goes back to early 1996.

\begin{figure*}[t]
\psfig{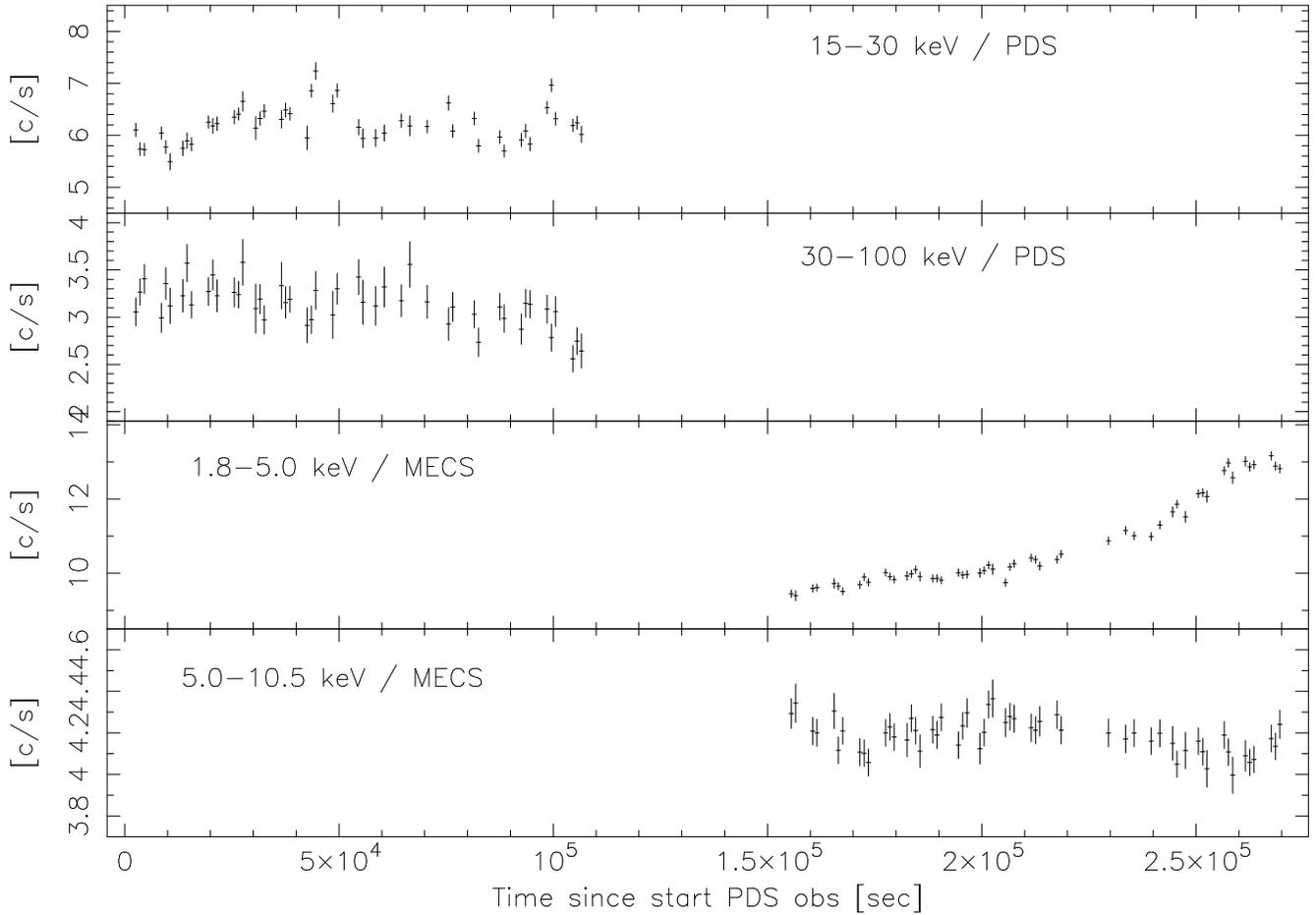}
\caption{History of observed intensities in four energy bands with a time
resolution of 1000~s. The background is negligible in the MECS data
for this bright source and was not subtracted from the relevant data
points. The time span covers 2002 March 31.7 to April 3.8 UT.
\label{fignfilc}}
\end{figure*}

The ASM light curves hint at quick spectral changes during the
outburst. The first transition started on March 31 (MJD~52364) and
lasted four days during which the 5-12 keV to 1.5--5 keV hardness
ratio dropped from about 6 to 2. A second transition started on April
16 (MJD~52380) and also lasted four days when the ratio dropped
further from 1.9 to 1.2, but now the drop is due to a decrease of the
5-12 keV flux rather than a 2--5 keV flux increase like in the first
transition.

\subsection{NFI}

The discovery prompted a target-of-opportunity observation (TOO) by
BeppoSAX. Since operations of BeppoSAX were seriously affected in
January 2002 when one of the two batteries experienced a partial
malfunction, power constraints prohibited the use of more than one
Narrow-Field Instrument at the same time. Therefore, two observations
were carried out on \bron: the first on March 31.7 through April 1.9
with the PDS (net exposure time 44.9 ksec) and the second on April 2.4
through 3.8 with the MECS (56.6 ksec). The latter observation was the
basis for the above-mentioned accurate X-ray localization.  The timing
of the observations was fortunate, because they cover the first
spectral state transition as seen with the ASM.

The MECS operates between 1.8 and 10.5 keV and is an imaging device
with a field of view of about 50\arcmin\ diameter with few-arcmin
resolution (Boella et al. 1997b). Fig.~\ref{figmecsimage} shows the
full-bandpass image obtained from the MECS data. It shows three
sources. Apart from \bron, they are the accretion-powered X-ray pulsar
4U~1907+097 (for a recent paper, see Baykal et al. 2001) and the soft
gamma-ray repeater and rotation-powered pulsar SGR~1900+14 (e.g.,
Kouveliotou et al. 1999). 4U~1907+097 has a pulse period of 441~s
which is clearly detectable. The photon count rate of SGR~1900+14 is
too small for a significant detection of its 5.16~s pulse period. MECS
data on \bron\ were extracted within 4\arcmin\ from the source
centroid. The background was obtained from long independent
measurements near the galactic poles, and was checked from an offset
position in the present observation. Spectra were binned to oversample
the resolution (8\% FWHM at 6~keV) by a factor of 3, and to obtain at
least 20 photons per bin to validate the use of the $\chi^2$
statistic. A 1\% systematic error per bin was assumed to allow for
small remaining calibration uncertainties.

The PDS acts between 15 and 250 keV, is a non-imaging device and
consists of 4 units that pair-wise rock between an on-source pointing
and two off-source pointings on opposite sides of the on-source
pointing at a distance of 2\fdg5 (Frontera et al. 1997). Each unit has
a collimator that restricts the field of view to 1\fdg3 (FWHM). Apart
from \bron, the field of view of the on-source pointing contains two
other bright X-ray sources that we are aware of, namely those detected
in the MECS image. We note that the {\em Chandra\/} observation does
not show any other bright source within 5\arcmin. The field of views
of the two off-source pointings contain no bright X-ray sources.

Figure~\ref{fignfilc} shows the time profiles in two energy bands of
the background-subtracted photon rate as determined from the PDS data
and in two bands as determined from the MECS data. There are trends on
long time scales that differ strongly from band to band. The largest
change in the flux occurs below 5 keV (3rd panel from the top) which
increases 40\% and appears to reach a maximum just before the end of
the observation. The rise accelerates at 2.2$\times10^5$~s after the
start of the PDS observation. This change at low energies must be
related to the first spectral transition seen in the ASM data.  The
PDS is likely contaminated by the same two sources that are visible in
the MECS data.  Fortunately, these have a unique signature which make
them easy to identify in the PDS data: both are pulsars. The PDS data
show a clear 441~s signal from 4U~1907+097 below 30 keV. The 5.16~s
pulsar in SGR~1900+14 is not detectable. Therefore, we ignore
$<30$~keV data as well as the variability in that band
(Fig.~\ref{fignfilc}, 1st panel). We note that no other remarkable
features, such as bursts or dips, were detected in the light curve (up
to a time resolution of 1~sec).

We have performed a period search of the MECS data from 0.2 to
10$^4$~s and on the whole bandpass, as well as below and above 5 keV,
and find no evidence for coherent oscillations with a 3$\sigma$ upper
limit to the amplitude of 1\% for a sinusoidal pulse profile. No
pulsations were discovered in the PDS data as well, with an upper
limit about 6 times as high as for the MECS data.

We also analyzed the PDS data that were reported by Feroci \& Reboa
(2002). This pertains to an exposure time of 96.6 ksec for both PDS
units combined, taken between March 9.4 and 12.2. The target was 
SGR~1900+14, so that \bron\ is at a 24\arcmin\ off-axis position which
implies a vignetting factor of 0.69. 4U~1907+097 is also in the field
of view at an off-axis angle of 47\arcmin. No bursts were detected.

In the following all quoted errors are the parameter ranges for which
$\chi^2<\chi^2_{\rm min}+2.71$ which is equivalent to 90\% confidence
for a single parameter.

\subsection{WFC}

The BeppoSAX Wide Field Cameras (WFCs; Jager et al. 1997) measured
\bron\ during two short exposures in the final days of operation. The
first was from April 27.6 to 27.8 UT for a net exposure time of 6.7
ksec and the second from April 29.4 to 29.7 for 8.1 ksec. Both were
with WFC unit 2. No variability was detected.

\section{Broad band spectrum}
\label{broad}

\begin{figure}[t]
\psfig{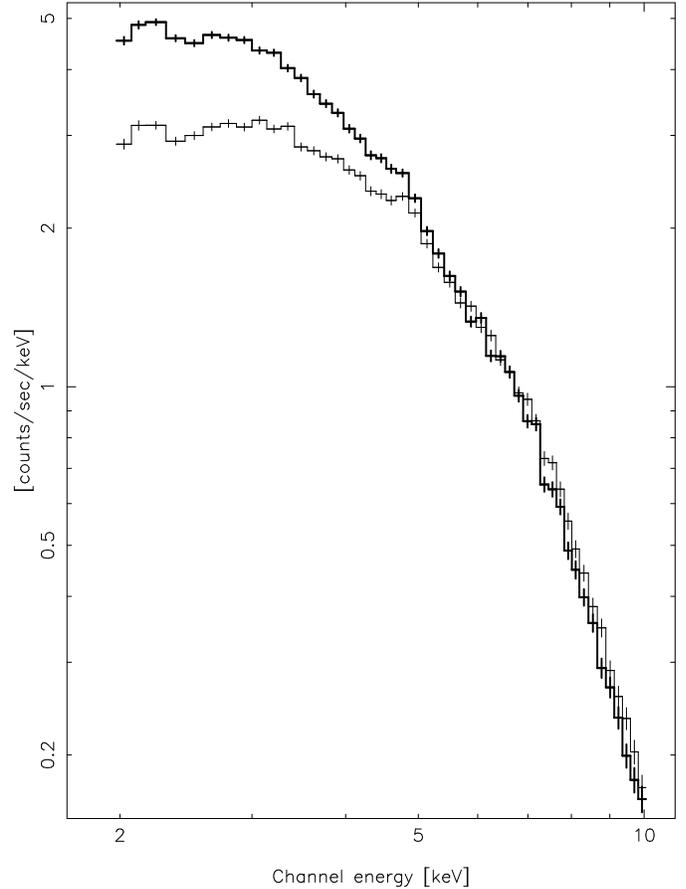}
\caption{The observed MECS spectra of the first (thin line) and last
intervals (thick line)
\label{fig2spectra}}
\end{figure}

\begin{figure}[t]
\psfig{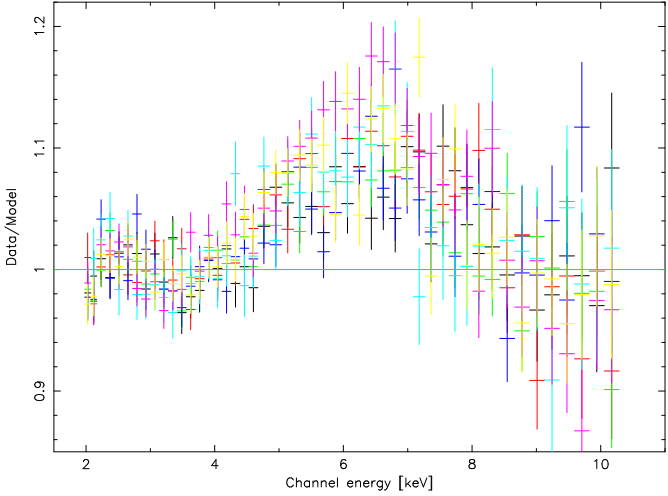}
\caption{Ratio of the observed spectrum to the predicted spectrum on
the basis of an absorbed power law plus multicolor accretion disk
black body as fitted to the 1.8--4.0 and 8.0-10.5 keV ranges.
\label{figratio}}
\end{figure}

\begin{figure}[t]
\psfig{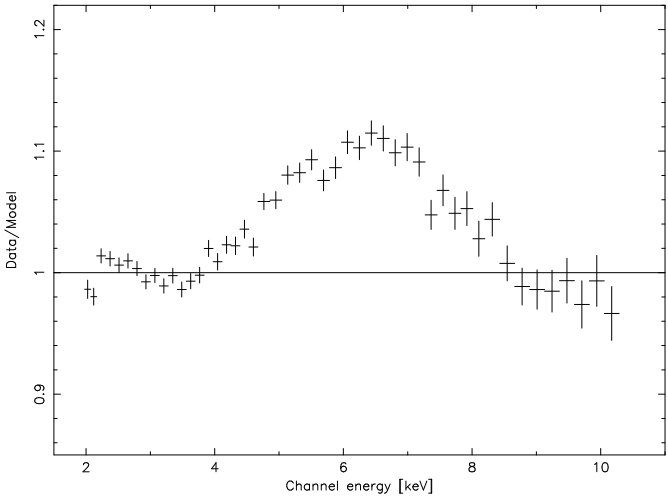}
\caption{Average ratio of the observed spectrum to the predicted spectrum.
\label{figratio2}}
\end{figure}

\begin{figure}[t]
\psfig{figure=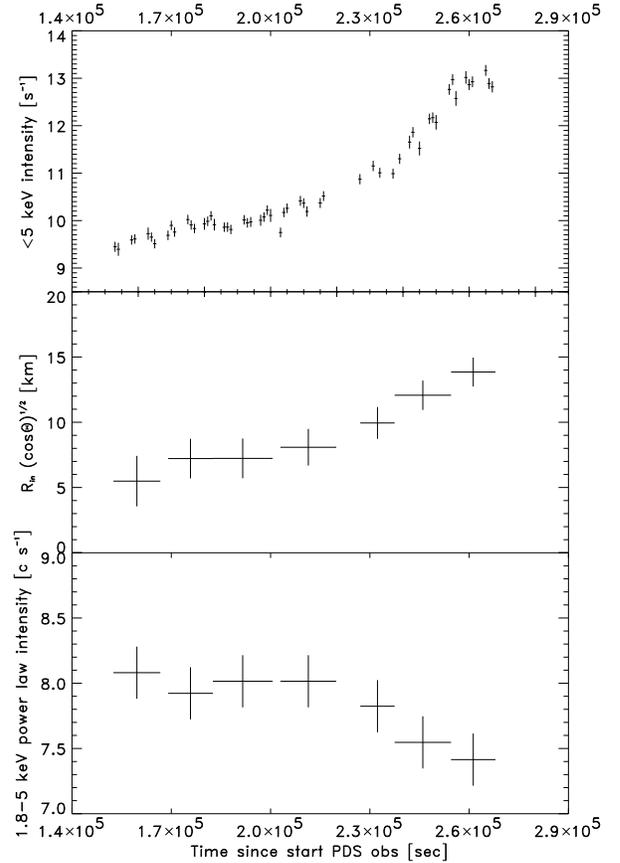,width=\columnwidth,clip=t}
\caption{Best-fit parameter values during the MECS observation. The
upper panel repeats the $<5$~keV light curve for guiding purposes.
\label{figfits}}
\end{figure}

In order to understand the nature of the energy-dependent flux
changes, we divided the MECS data in seven time intervals in which the
source shows a particular state in the 2--5 keV data. The time spans are
between 15 and 20~ksec. Figure~\ref{fig2spectra} illustrates the
spectral variability by showing the spectra for the first and seventh
interval. This, again, shows strong changes below about 5 keV and weak
and unrelated changes above that. The first spectrum is fairly well
described by a simple absorbed power law ($\chi^2/\nu=1.435$,
$\nu=45$) with a hydrogen column density of $N_{\rm
H}=(2.34\pm0.07)\times10^{22}$~cm$^{-2}$ (assuming the model by
Morrison \& McCammon 1983) and a photon index of
$\Gamma=1.89\pm0.02$. This model does not fit the last spectrum
($\chi^2/\nu=2.772$, $\nu=45$). Inspection of the fit residuals
suggests the presence of an extra component between 2 and 4 keV which
is obvious from Fig.~\ref{fig2spectra}. When modeling this by a
multicolor accretion disk black body (Mitsuda et al. 1984) the fit
becomes satisfactory ($\chi^2/\nu=1.16$, $\nu=43$). Therefore, we
simultaneously fitted all seven spectra (335 spectral bins) with an
absorbed power law plus a multicolor disk black body, leaving free
over all spectra the black body and power law normalizations, while
coupling $N_{\rm H}$, the accretion disk inner edge temperature
k$T_{\rm in}$ and the power law index. Since k$T_{\rm in}$ is well
below the lower threshold of the bandpass (and the spectrum maximum
just at that threshold), there is a strong dependence between this and
the black body normalization and we refrain from varying the
temperature. The resulting fit is inadequate ($\chi^2/\nu=1.796$,
$\nu=318$). The fit residuals show a noisy excess between 4 and 7 keV
which is illustrated in Figs.~\ref{figratio} and
\ref{figratio2}. Therefore, we included a Gaussian function in the
model that is constant over all seven spectra.  The improvement in
$\chi^2/\nu$ is dramatic, down to 1.020 ($\nu=315$). For this fit,
$N_{\rm H}=(2.50\pm0.16)\times10^{22}$~cm$^{-2}$ and k$T_{\rm
in}=0.77\pm0.03$~keV. In Fig.~\ref{figfits} are shown the fitted
values for the black body normalization and the photon count rate in
the MECS below 5 keV as predicted for the power-law component.  It is
clear that the increase in flux is fully explained by the black body
component, while the power law component remains rather stable. We
note that the ASM measurements show an increase in $>5$~keV flux
as well which must have occurred shortly after the end of the
MECS observation.

\begin{figure}[t]
\psfig{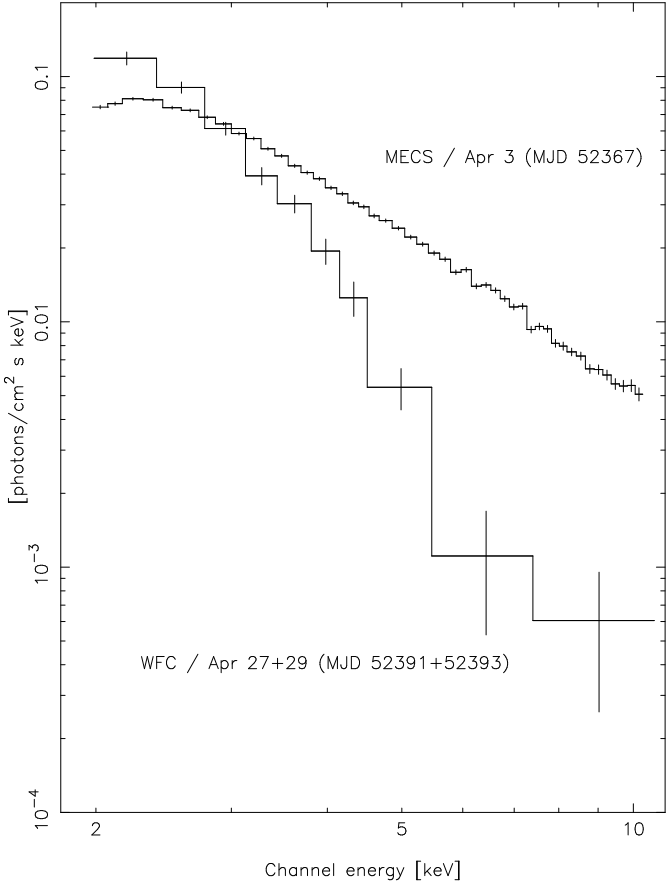}
\caption{Incident photon spectra as observed by the MECS (last
interval) and the WFC (both WFC observations combined), exhibiting a
strong softening of the source.
\label{figwfcmecs}}
\end{figure}

\begin{table}[tb]
\caption[]{Parameter values of the best-fit spectral model to the MECS
data (335 spectral bins distributed over 7 spectra) and the two PDS data
sets. Errors are 90\% confidence per parameter of interest (i.e., from
range in each parameter for which $\chi^2$ is smaller than minimum
$\chi^2$ plus 2.71).\label{tab1}}
\begin{tabular}{ll}
\hline\hline
\multicolumn{2}{c}{MECS data} \\
\hline
Model$^\ddag$        & {\tt wa$\times$(po+diskbb+gauss)} \\
$N_{\rm H}$          & $2.50\pm0.16~10^{22}$~cm$^{-2}$ \\
k$T_{\rm in}$        & 0.77$\pm0.03$~keV \\
$\Gamma$             & $1.91\pm0.08$ \\
Gauss $E_{\rm line}$ & 5.90$\pm0.60$ keV\\
Gauss line width     & 3.2$\pm0.5$ keV (FWHM) \\
Gauss line flux      & $(6.9\pm4.6)~10^{-3}$~phot~s$^{-1}$cm$^{-2}$ \\
Gauss line EW        & 0.27$\pm0.13$ keV \\
Unabs. 2-10 keV flux & from 1.49 to 1.76$\times10^{-9}$~\ecs\\
$\chi^2/\nu$         & 1.020 (315 dof, 1\% systematic error) \\
\hline
Model$^\ddag$        & {\tt wa$\times$(po+diskbb+laor)} \\
$N_{\rm H}$          & $2.46\pm0.05~10^{22}$~cm$^{-2}$ \\
k$T_{\rm in}$        & 0.77$\pm0.03$~keV \\
$\Gamma$             & $1.95\pm0.03$ \\
Laor $E_{\rm line}$  & 6.4$^{+0.03}$ keV\\
Laor $R_{\rm in}$    & 3.9$^{+1.6}_{-0.6}$ G$M_{\rm compact}$/c$^2$ \\
Laor inclination     & $90_{-3.8}$~deg \\
Laor line flux       & $(3.0\pm0.4)~10^{-3}$~phot~s$^{-1}$cm$^{-2}$ \\
Laor line EW         & 0.23$\pm0.12$~keV \\
Unabs. 2-10 keV flux & from 1.49 to 1.77$\times10^{-9}$~\ecs\\
$\chi^2/\nu$         & 1.108 (314 dof, 1\% systematic error) \\
\hline\hline
\multicolumn{2}{c}{30--250 keV PDS data March 31} \\
\hline
Model$^\ddag$        & {\tt po$\times$highecut} \\
$\Gamma$             & 2.10$\pm0.05$ \\
$E_{\rm c}$          & 65$\pm7$ keV\\
$E_{\rm f}$          & 158$\pm28$~keV\\
30-250 keV flux      & 2.1$\times10^{-9}$~\ecs\\
$\chi^2/\nu$         & 0.750 (49 dof) \\
\hline\hline
\multicolumn{2}{c}{30--250 keV PDS data March 9} \\
\hline
Model$^\ddag$        & {\tt po$\times$highecut} \\
$\Gamma$             & 1.90$\pm0.04$ \\
$E_{\rm c}$          & 66$\pm4$ keV\\
$E_{\rm f}$          & 143$\pm16$~keV\\
30-250 keV flux      & 3.2$\times10^{-9}$~\ecs\\
$\chi^2/\nu$         & 1.05 (48 dof) \\
\hline\hline
\end{tabular}
\label{tabfits}
$^\ddag${\tt wa} - absorption following the model by Morrison \& McCammon
(1983); {\tt po} - power law with photon index $\Gamma$; {\tt diskbb} -
multi-temperature disk black
body following Mitsuda et al. (1984); {\tt gauss} - Gauss line profile;
{\tt laor} - relativistically broadened line profile following Laor (1991);
{\tt highecut} - exponential cutoff function with e-folding cutoff energy
$E_{\rm f}$ above a certain threshold energy $E_{\rm c}$.

\end{table}

As an alternative to the Gaussian model for the 4-7 keV feature, we
applied a model for a relativistically broadened emission line near a
compact object following Laor (1991), leaving free the line energy
(but limiting the freedom between 6.4 and 6.97 keV), the radius (in
terms of Schwarzschild radii) of the inner edge of the accretion disk,
the inclination angle and the flux. The fit is acceptable with
$\chi^2/\nu=1.108$ ($\nu=314$), and the fit parameters are given in
Table~\ref{tab1}. The inclination angle is high, testifying to the
symmetry of the feature and the fact that a Gaussian line fits just as
good or slightly better.

We tested for the presence of an absorption edge between 7.1 and 9.3
keV, as would be expected for the iron K-shell at these energies, but
an f-test predicts a 36\% chance probability for the observed
improvement in the fit.

The 30-250 keV PDS data taken immediately before the MECS data are
inconsistent with a single power law ($\chi^2/\nu=3.308$,
$\nu=51$). If an exponential cutoff is introduced, the fit becomes
acceptable with $\chi^2/\nu=0.750$ ($\nu=49$) and the photon index of
$2.10\pm0.05$ is consistent with the MECS data. The fitted values are
listed in Table~\ref{tab1}.

We performed a simultaneous fit to the first MECS spectrum (in
1.8--10.5 keV) and the data from the final 15\% of the PDS observation
(30--250 keV). These two spectra were taken about one day apart. The
fit with the absorbed cut-off power law plus disk black body and
gaussian line is quite good ($\chi^2/\nu=0.943$ with $\nu=90$) and the
normalization of the PDS to the MECS data is within limits
($1.16\pm0.14$). The parameter values are consistent with those
obtained when treating the data independently (see
above). Alternatively, we fitted this 2--250 keV spectrum with a
Comptonized spectrum (following Titarchuk 1994 whereby the soft
photons that are up scattered are described by a Wien spectrum with a
temperature below our bandpass) plus multicolor accretion disk black
body. The fit is good as well ($\chi^2/\nu=0.886$ with $\nu=90$). The
fitted parameters are: $N_{\rm H}=(2.4\pm0.5)\times10^{22}$~cm$^{-2}$,
plasma temperature $39\pm14$~keV, optical depth $1.1\pm0.1$ (for a
disk geometry) or $3.2\pm0.3$ (for a spherical geometry), and the PDS
to MECS normalization is again within limits ($1.19\pm0.20$). The
1.8-250 keV unabsorbed flux is 4.6$\times$10$^{-9}$~\ecs. If the disk
component extends to lower energies, the flux could be a factor of 2
higher. The 1.8-10 keV unabsorbed flux increases by
0.3$\times$10$^{-9}$~\ecs\ during the course of the MECS observation.

The PDS spectrum taken on March 9.4--12.2 by Feroci \& Reboa (2002)
looks quite similar above 30~keV to that taken between March 31.7 and
April 1.9 (Table~\ref{tab1}). The same cutoff law applies while the
photon index is slightly harder (1.9 versus 2.1).

We combined the WFC spectral data from the observations on April 27
and 29, after verifying that the source was stable between both
dates. The total exposure time is 14.9 ksec. The spectrum is much
softer than during the MECS observation 24 days earlier, see
Fig.~\ref{figwfcmecs}.  An absorbed power law with $N_{\rm
H}=2.5\times10^{22}$~cm$^{-2}$ fixed provides an adequate description
of the data ($\chi^2/\nu=1.35$, $\nu=26$) with $\Gamma=4.5\pm0.2$~keV,
but a disk black body fits better with $\chi^2/\nu=0.68$ ($\nu=26$)
and k$T_{\rm in}=0.62\pm0.03$ and $R_{\rm
in}(cos\theta)^{1/2}=39\pm6$~km (for a distance of 10~kpc). The
unabsorbed 2--10 keV flux is 1.1$\times10^{-9}$~\ecs, or 30\% less
than measured during the first MECS spectrum. However, compared with
the black body component in the last MECS spectrum only, the flux is
three times higher. There is no significant detection of a power-law
component; it is at least a factor of 5 fainter than during the MECS
observation.

\section{Discussion}
\label{discussion}

Woods et al. (2002) proposed a BH hypothesis for the compact object in
\bron, on the basis of a hard spectrum similar to other accreting BHs
in the low/hard state (for a review, see Tanaka \& Shibazaki 1996) and
the lack of pulsations.  Our observations support this proposal. We
have analyzed a total of $\sim210$~ksec exposure on \bron\ (excluding
the ASM coverage), 2/3 of it above 15~keV, and fail to detect
signatures of a neutron star: there are pulsations nor type-I X-ray
bursts. Definitive proof for the BH nature needs to come from radial
velocity measurements of the near-infrared counterpart identified by
Chaty \& Mignani (2002).

With this source, the total number of X-ray binaries suspected to
inhabit a BH is about 31. Half of these have been dynamically
confirmed (J.~Orosz, priv. comm.) through a mass function larger than
3~M$_\odot$.

The 2--250 keV spectrum is characterized by two continuum components:
a multicolor accretion disk black body and a Comptonization spectrum.
The black body component is on the rise during the MECS observation,
when its flux triples within the bandpass. Since a major part of the
black body is outside our bandpass (k$T_{\rm in}\approx$0.8 keV) and
the interstellar absorption is high (implying $\approx$90\% absorption
at 2~keV), it is impossible to study the evolution of its temperature
and geometric size. However, it is reasonable to assume that the rise
is associated with the inner edge of the disk getting closer to the
compact object. During the WFC observation 24 days later, the 2--28
keV spectrum is dominated by this component but the temperature is
still below 1~keV. The spectral evolution of the outburst is
concentrated in two fairly discrete softening events around April 2
and 16. The first event is what was observed with the MECS. Here, the
softening is mainly caused by a brightening of the black body
component. The second event appears to be due to a strong decay of the
Comptonization, as suggested by the ASM data. Perhaps the plasma is
effectively cooled down by the black body photons at this point. The
two spectral state transitions are probably related to changes in the
mass accretion rate and another parameter, and are similar to what is
observed in other (BH) transients (e.g., Nowak 1995; Tanaka \&
Shibazaki 1996; Belloni 2001; Homan et al. 2001): initially, the
source was in the 'low/hard state' ($\Gamma\sim1.55$) with a high
fractional variability of 43\% rms (Woods et al. 2002). The first
transition is probably into the 'intermediate state' when a disk black
body appears while the power law remains present. Two weeks later the
source appears to transit into the 'high/soft state' when the
Comptonized component drops significantly (as evidenced by the WFC
observation) and the disk black body is considerably brighter. The
source did not move to the 'very high state' for which the disk
temperature should increase to the 1-2 keV regime. More diagnostics,
particularly pertaining to the timing behavior, may be deduced from
(mostly private) PCA data.

In addition to the two continua components there is an emission
feature centered on the location of the Fe-K complex which is very
broad (3.2$\pm0.5$~keV FWHM). Similar features have been detected in
other X-ray binaries, usually but not exclusively (e.g., Oosterbroek
et al.  2001) in BH transients. A few recent examples of BH cases are
XTE J1650-500 (Miller et al. 2002), SAX J1711.6-3808 (In 't Zand et
al. 2002b), XTE J2012+381 (Campana et al. 2002) and GRS~1915+105
(Martocchia et al. 2002). The widths are similar. Generally, two
explanations are proposed for the broadening: Compton scattering in a
corona or relativistic Doppler broadening in combination with
gravitational redshift in a part of the accretion disk close to the
black hole. As we have determined in our spectral analysis,
relativistic Doppler broadening provides a good description of the
line in \bron. However, the same applies for Compton broadening in a
$kT\approx40$~keV/$\tau=1.1$ plasma: following Czerny et al. (1991)
one expects a line width of about 3.1~keV (FWHM).

A substantial part of the bolometric flux is outside the 2--10 keV
band. During the MECS+PDS observation on March 31 through April 3, the
10--250 keV flux is roughly twice the 2--10 keV flux. Furthermore,
there may be just as much flux below 2~keV as above and we estimate
that the bolometric flux is about 1$\times10^{-8}$~\ecs. Another
measurement of the broad-band flux can be obtained if we combine the
PCA observation on March 17 by Woods et al. (2002), which translates
to an unabsorbed 2--30 keV flux of $4\times10^{-9}$~\ecs, and the PDS
measurement on March 9--12 of 3.2$\times10^{-9}$~\ecs in 30--250
keV. The ASM light curve suggests that no substantial flux changes
happened between both observations. The summed 2--250 keV flux is
7$\times10^{-9}$~\ecs\ which is 50\% above the PDS+MECS measurement 3
weeks later. Since there are only two measurements above a few tens of
keV and none possible below 2~keV due to the high interstellar
absorption, it is uncertain what the peak bolometric flux is. Also,
ASM measurements show an increase by a factor of two at energies above
5~keV soon after the MECS+PDS measurement. This could be due to the
tail of the brightening black body component, or to an increase of the
power-law component. Either way, we estimate that the 2-250~keV flux
may be about $1.0\times10^{-8}$~\ecs\ during the peak as timed by the
ASM light curve. Taking into account the invisible flux below 2~keV,
we estimate that the peak bolometric flux may have been roughly
$2\times10^{-8}$~\ecs. The state transitions indicate that the source
reached luminosities in excess of $\approx$10\% of the Eddington limit
(e.g., Nowak 1995) which is 2$\times$10$^{37}$~\lum\ if we assume that
the mass is greater than that of a neutron star. This yields a lower
limit to the distance of 3~kpc (6~kpc if we are dealing with a BH with
a minimum mass of 3~M$_\odot$).

\acknowledgement We thank the BeppoSAX Mission Scientist Luigi Piro
and the Time Allocation Committee for granting considerable amounts of
Project Time to carry out the March 31 TOO, and the Mission Planning
team for their enthusiastic effort to schedule it under many
constraints. We are grateful to Marco Feroci for generously providing
the PDS March 9 data on SGR~1900+14. JZ acknowledges financial support
from the Netherlands Org. for Scientific Research (NWO).

\end{document}